\documentclass[conference]{IEEEtran}
\usepackage[bookmarks=true,breaklinks=true,colorlinks,linkcolor=blue,citecolor=blue,urlcolor=blue]{hyperref}
\usepackage{cite}
\usepackage{booktabs}
\usepackage{graphicx}
\usepackage{multirow}
\usepackage{enumitem}
\usepackage{wrapfig}

\newcommand{\taskOne}{Amoeba}

\begin{document}

\title{System Support for Environmentally Sustainable Computing in Data Centers\vspace{-8pt}}

\author{
Fan Chen\\
Luddy School of Informatics, Computing, and Engineering, Indiana University, Bloomington, IN\\
E-mail: fc7@iu.edu\vspace{-16pt}\\
}

\maketitle

\begin{abstract}
Modern data centers suffer from a growing carbon footprint due to insufficient support for environmental sustainability. While hardware accelerators and renewable energy have been utilized to enhance sustainability, addressing Quality of Service (QoS) degradation caused by renewable energy supply and hardware recycling remains challenging:
(1) prior accelerators exhibit significant carbon footprints due to limited reconfigurability and inability to adapt to renewable energy fluctuations;
(2) integrating recycled NAND flash chips in data centers poses challenges due to their short lifetime, increasing energy consumption;
(3) the absence of a sustainability estimator impedes data centers and users in evaluating and improving their environmental impact.
This study aims to improve system support for environmentally sustainable data centers by proposing a reconfigurable hardware accelerator for intensive computing primitives and developing a fractional NAND flash cell to extend the lifetime of recycled flash chips while supporting graceful capacity degradation. We also introduce a sustainability estimator to evaluate user task energy consumption and promote sustainable practices. 
We present our preliminary results and recognize this as an ongoing initiative with significant potential to advance environmentally sustainable computing in data centers and stimulate further exploration in this critical research domain.
\vspace{-4pt}
\end{abstract}


\vspace{-4pt}
\section{Introduction}
Modern data centers contribute significantly to an unsustainable carbon footprint, from both operational and embodied sources. \textit{Operational} emissions arise from routine hardware usage, while \textit{embodied} emissions result from manufacturing and construction processes. Global data center operational energy is projected to reach 4.5\% of global demand by 2025~\cite{Liu:GEI2020}. Additionally, Over the past decade, global data center infrastructure capacity, reflecting embodied carbon footprint, has surged $6\times$~\cite{Masanet:Science2020}. 
\textit{Renewable energy} and \textit{hardware recycling} are two critical methods to enhance the environmental sustainability of future data centers. Several cloud companies~\cite{Acun:ASPLOS2023} have collectively invested in 22 GW of renewable energy generation to reduce the operational carbon footprint.  Meanwhile, recent work considers reusing obsolete hardware, e.g., mobile CPUs~\cite{Switzer:ASPLOS2023}, mobile DRAMs~\cite{Switzer:ASPLOS2023}, and FPGAs~\cite{DoganISDFT2014}, to reduce the embodied carbon footprint in data centers.
However, mitigating Quality of Service (QoS) degradation in future sustainable data centers caused by renewable energy supply and hardware recycling poses a complex challenge.

On one side, renewable energy supply significantly degrades QoS due to its intermittent nature. Solar and wind generate 47\% and 34\% of the renewable energy~\cite{Nalley:EIA2022}, leading to hourly and seasonal fluctuations in energy supply. While large energy storage~\cite{Acun:ASPLOS2023,Singh:NSDI2013} can mitigate these fluctuations, their addition significantly increases the embodied carbon footprint of data centers. Consequently, servers may need to be slowed down or shut down during low energy supply periods.
Software-based techniques such as carbon-aware scheduling~\cite{Acun:ASPLOS2023} and virtual transient servers~\cite{Singh:NSDI2013} are proposed to schedule non-critical tasks among CPUs without significant QoS degradation in the face of energy supply fluctuations. However, emerging workloads such as big data, deep learning, and cryptography heavily use accelerators~\cite{Chen:MICRO2014,Shafiee:ISCA2016} and large data storage~\cite{Grupp:FAST2012}. \textit{There is a near total lack of hardware (particularly accelerator and storage) support for maintaining the QoS in data centers powered by renewable energy supply}.

On the other side, while hardware recycling reduces the embodied carbon footprint, it compromises data center QoS due to the limited performance and reliability of obsolete hardware. Recent attempts to integrate outdated computing components like mobile processors~\cite{Switzer:ASPLOS2023} and DRAMs~\cite{Switzer:ASPLOS2023} fall short in handling demanding tasks in emerging workloads. Thus, future data centers will require reconfigurable application-specific hardware acceleration with low embodied carbon footprints. 
Although few works~\cite {Boyd:ITSM2011} explore reusing obsolete NAND flash chips in data center storage systems, utilizing aging flash chips with about-to-worn-out blocks poses challenges in supporting long-term workloads in future data centers.

Below, we provide a detailed overview of these challenges across the computing and storage layers of system stacks:
\begin{itemize}[leftmargin=*, topsep=0pt, partopsep=0pt, itemsep=0pt]
\item \textbf{Hardware Acceleration}: Previous GPU-, FPGA-~\cite{DoganISDFT2014,Isaka:TDMR2022}, ASIC-~\cite{Chen:MICRO2014} and PIM-based~\cite{Shafiee:ISCA2016} accelerators have been utilized to mitigate  performance loss caused by hardware recycling. However, none effectively addresses energy supply fluctuations or maintains a low embodied carbon footprint. Specifically, GPU- and application-specific-integrated-circuit (ASIC)-based accelerators lack the ability for rapid wake-up or shutdown in response to energy supply fluctuations. 
While FPGA-~\cite{Liauw:ISSCC2012,Chen:ISLPED2010} and PIM-based~\cite{Shafiee:ISCA2016,Qiu:HPCA2020} accelerators adopt nonvolatile memory arrays for data retention, critical components such as SRAM-based switches in FPGAs and analog-to-digital converters (ADCs) in PIMs still lose information without power, hindering prompt workload resumption after power loss.  
ASIC-based accelerators~\cite{Chen:MICRO2014} focusing on accelerating only one type of computing kernels suffer from high embodied carbon footprint, due to their lack of reconfigurability. 

\item \textbf{NAND Flash}: To reduce embodied carbon footprint, old NAND flash chips will be adopted by future sustainable data centers. However, flash suffers from short cell endurance~\cite{Yu:DATE2012}, where \textit{about-to-worn-out} blocks limit chip lifetime. Despite error correcting code (ECC)~\cite{Nakamura:JSSCC2019,Tanzawa:JSSCC1997} and wear leveling~\cite{Jimenez:USENIX2014} assistance, no prior technique significantly extends the lifetime of about-to-worn-out blocks, hindering widespread adoption of recycled NAND flash chips in future data centers. Moreover, due to fluctuations in renewable energy supply, DRAM snapshots are frequently stored in and loaded from storage systems. Unfortunately, most recycled flash chips use multi-level cells (MLCs)~\cite{Yu:ICCD2013}, suffering from long program and read operations.

\item \textbf{Sustainability Estimator}: Without an environmental sustainability estimator, data centers struggle to assess the operational and embodied energy of user workloads, leading to a lack of visibility regarding the sustainability costs of user actions. Additionally, the absence of such metrics inhibits the development of flexible billing policies that could reward users who opt for recycled hardware and encourage broader adoption of environmentally friendly practices.
\end{itemize}

\vspace{-0pt}
\section{Approaches}
\vspace{-2pt}
This project aims to develop system support for environmentally sustainable data centers, with three main objectives:
(1) Facilitate energy-efficient hardware acceleration for memory-intensive tasks such as big data, deep learning, and cryptography, ensuring steady progress even in the face of renewable energy fluctuations;
(2) Extend the lifetime of about-to-worn-out blocks in recycled NAND flash chips, thereby mitigating the embodied energy consumption;
and (3) Provide an accurate evaluation of both operational and embodied energy consumption for data center user tasks and incentivize users to make their computing more environmentally sustainable.
The primary approach entails constructing an environmentally sustainable computing system.
This system encompasses a nonvolatile reconfigurable PIM accelerator, a storage system based on recycled NAND flash, and an accurate environmental sustainability estimator.

\vspace{-2pt}
\subsection{\taskOne~-~A Reconfigurable Nonvolatile Accelerator}
\begin{wrapfigure}{b}{0.25\textwidth}
\begin{center}
\vspace{-6pt}
\includegraphics[width=\linewidth]{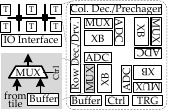}
\end{center}
\vspace{-16pt}
\caption{The {\taskOne}~architecture.}
\label{f:task1_overview}
\vspace{-10pt}
\end{wrapfigure}

\vspace{-2pt}
Existing PIMs in data centers encounter substantial manufacturing costs and operational inefficiencies, particularly when faced with intermittent renewable energy sources, resulting in significant embodied and operational carbon emissions. 
We propose \textit{\taskOne}, depicted in Figure~\ref{f:task1_overview}, as a reconfigurable FeFET-based PIM architecture aimed at addressing these challenges.
Our approach involves profiling compute- and resource-intensive computational kernels in data centers, including NTT for lattice-based cryptography, SHA3 for blockchains, and convolution for machine learning, leveraging insights from prior studies.
Categorizing basic operations and supporting computation primitives with three types of processing engines (PE) based on the FeFET crossbar structure, \taskOne~offers fine-grained PE-level reconfiguration to markedly minimize embodied carbon emissions. 
Furthermore, serving as a fully nonvolatile accelerator, it ensures consistent forward progress even amidst renewable energy fluctuations, thereby contributing to a notable reduction in the operational carbon footprint.

\textbf{PE-Level Reconfigurability for Reducing Embodied Carbon Footprint}. Crossbars in \taskOne~can be logically configured into different modes:
(1) \textbf{Associative PE (APE) for \texttt{LUT} and \texttt{ADD}}.
Each crossbar acts as an embedded CAM block for associative computing.  As in previous work~\cite{ASPDAC2020PARC}, each unit consists of complementary cells (\texttt{H} and \texttt{L}) for 1' and 0', respectively. The search word is encoded by complementary voltages (\texttt{V}/\texttt{0} for 1' and \texttt{0}/\texttt{V} for 0') and broadcasted onto vertical search lines. After a parallel search, the horizontal match line identifies matches and outputs the vector for post-processing. \texttt{LUT} is directly implemented. For \texttt{ADD}, bitwise search-based additions are cascaded for multi-bit addition, allowing for parallel execution.
(2) \textbf{Multiplicaiton PE (MPE) for \texttt{MVM} and \texttt{SHIFT}}.
The crossbar-based \texttt{MVM} unit, as in~\cite{ASPDAC18gan, DAC2019}, maps dataflow in a weight-stationary manner. Weights are pre-coded onto columns, and input feature maps are converted to read voltages and applied to horizontal wordlines. The ADC aggregates current at vertical bitlines, representing dot-products.
\texttt{SHIFT} operations comprise $>$40\% of NTT workloads, but they aren't inherently \texttt{MVM} operations and can't be directly mapped to crossbars. We propose transferring \texttt{SHIFT} to \texttt{MVM} by pre-coding the transformation matrix onto the crossbar and applying the input as a multiplicand. This scheme, based on linear algebra, can be generalized to support any cyclic permutation.
(3) \textbf{Computing PE (CPE) for \texttt{LOGIC} operations}.
We perform \texttt{Logic} operations like \texttt{AND} and \texttt{XOR} within the crossbar array without using CMOS ALUs. In contrast to prior methods~\cite{zha2016reconfigurable, zha2018liquid}, which necessitate a larger crossbar array, our approach only requires a 2$\times$N crossbar, where $N$ represents the input width. Take \texttt{ADD} as an example, we adjust ADC sensing levels to output a logical '1' only when both cells are in a low-resistance state.
(4) \textbf{Combining APE and MPE for \texttt{MUL}}.
Prior algorithms~\cite{haj2018efficient, haj2018imaging} broke an N-bit \texttt{MUL} to \texttt{SHIFT} and \texttt{ADD} of partial products. \cite{nejatollahi2020cryptopim} implemented such \texttt{MUL} with bitwise operations based \texttt{ADD} and implicit data selection based \texttt{SHIFT}. Instead, we combine the APE and MPE to implement \texttt{MUL}. The benefits are twofold: first, the search-based \texttt{ADD} reduces expensive read/write operations in~\cite{nejatollahi2020cryptopim}; second, the \texttt{MVM} based explicit \texttt{SHIFT} reduces the complex control for memory block selection in~\cite{nejatollahi2020cryptopim}.  

\begin{figure*}[t!]
\centering
\includegraphics[width=.95\linewidth]{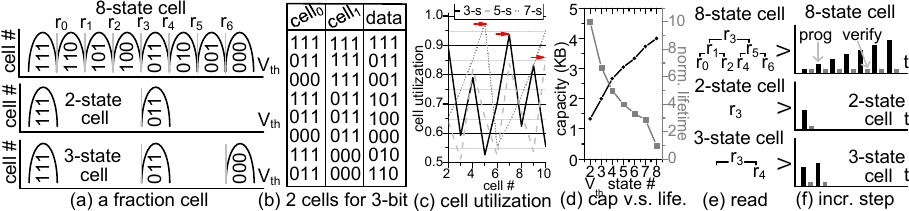}
\vspace{-0.15in}
\caption{A Fraction NAND flash cell (FRAC)}
\label{f:task_2_all}
\vspace{-0.26in}
\end{figure*}

\begin{figure}[h!]
\centering
\vspace{-12pt}
\includegraphics[width=\linewidth]{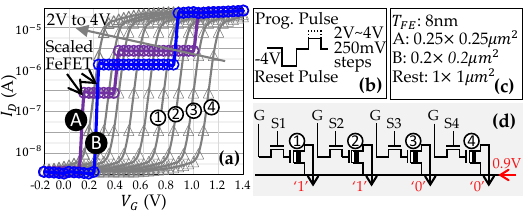}
\vspace{-24pt}
\caption{A fully nonvolatile accelerator.}
\label{f:task1_allnv}
\vspace{-14pt}
\end{figure}

\textbf{Fully Nonvolatile Accelerator for Reducing Operational Carbon Footprint}.
To reduce the \textit{operational} carbon footprint, we propose a fully nonvolatile accelerator with FeFET-based True Random Number Generator (TRG) and ultra-low power FeFET-based ADC. 
(1) \textbf{FeFET-based \texttt{TRG}}.
We leverage the stochastic switching behavior in scaled FeFET devices~\cite{deng2020comprehensive} as an on-chip entropy source for \texttt{TRG} designs, same as prior work~\cite{islped2021}. Our baseline \texttt{TRG} employs a 0.2$\mu{m}$$\times$0.2$\mu{m}$ FeFET device with a read voltage set at 0.2V. Initial findings indicate a bias towards `0's in the output bits. To mitigate this bias, we propose a tracking scheme involving an 8-bit counter to record output probabilities in consecutive 256-bit segments. The counter's output then acts as the control signal for adjusting write voltages in generating subsequent segments.
(2) \textbf{Low-Power ADC with Tunable Precision}.
We develop a precision-scalable, low-power ADC using FeFET technology for faster analog-to-digital conversion. Figure~\ref{f:task1_allnv}(d) illustrates a four-bit ADC. We utilize four partially polarized 1$\mu{m}$$\times$1$\mu{m}$ FeFET devices (\textcircled{1}$\sim$\textcircled{4}). The write line toggles between 0$V$ and $V_{DD}$ for enabling/disabling. Inputs are applied to the data line. Due to variations in $V_{th}$, read currents differ. 
For instance, with a 0.9V input in Figure~\ref{f:task1_allnv}(d), higher-than-threshold current is detected on \textcircled{1} and \textcircled{2}, while lower current is observed on \textcircled{3} and \textcircled{4}. This results in a 4-bit output "1100" with an appropriately set current threshold. The ADC's precision can be easily adjusted by disabling \textcircled{1} and \textcircled{3}. Threshold currents for each device can be dynamically programmed.

\begin{figure*}[t!]
\centering
\includegraphics[width=.95\linewidth]{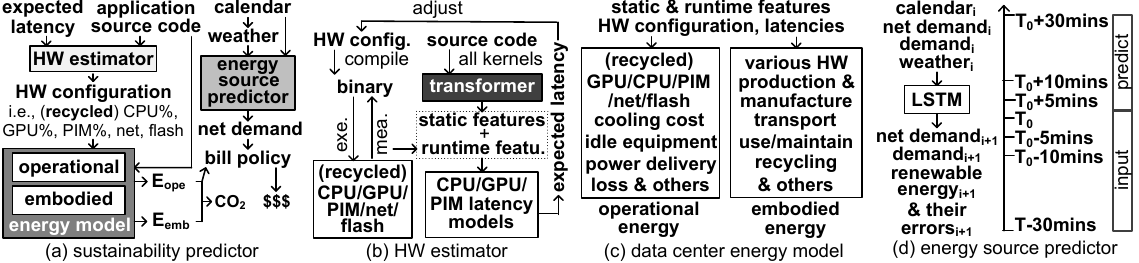}
\vspace{-0.15in}
\caption{An environmental sustainability estimator (ESE).}
\label{f:task_3_all}
\vspace{-0.2in}
\end{figure*}

\subsection{FRAC - Recycling Used NAND Flash Chips}
\vspace{-2pt}
To reduce embodied energy, repurposing outdated NAND flash chips in sustainable data centers is crucial. These chips, manufactured using older process technologies, may have undergone varying levels of write operations before recycling, leading to differing remaining lifetimes. 
While existing work~\cite{Wang:DATE2012,Hsieh:TECS2013,Lin:TVLSI2019} salvages already-worn-out blocks, the remaining lifetime of recycled chips is predominantly influenced by about-to-worn-out blocks. We propose FRAC to improve the lifetime of about-to-worn-out blocks by exploring the trade-off between chip capacity and cell lifetime. FRAC supports graceful storage capacity degradation by gradually decreasing the number of $V_{th}$ states in a cell. 

\textbf{FRAC: A NAND Flash Fraction Cell}.
A conventional flash cell (e.g., SLC, MLC, TLC, or QLC) uses $2^n$ $V_{th}$ states to store $n$ bits respectively, with $n\in[1,4]$.Figure~\ref{f:task_2_all}(a) illustrates a TLC cell.
In contrast, FRAC uses $m$ $V_{th}$ states to store data, where $m\in[2, 2^n]$, e.g., $m=3$. However, a fraction cell having 3 $V_{th}$ states can represent only 1 bit, wasting 1 $V_{th}$ state. To use more $V_{th}$ states in a cell, we propose to use $\alpha$ FRAC cells, each of which has $m$ $V_{th}$ states, to store $\lfloor\log_2(m^\alpha)\rfloor$ bits. For instance, two 3-state cells can store 3 bits. The truth table of two 3-state cells is shown in Figure~\ref{f:task_2_all}(b). By looking up this table, the value of two 3-state cells is translated to 3 bits. We define cell utilization as the number of $V_{th}$ states representing data divided by the total number of $V_{th}$ states of FRAC cells. We explore the cell utilization of 3-, 5-, and 7-state cells in Figure~\ref{f:task_2_all}(c). To achieve the highest cell utilization, we can store 11 bits in seven 3-state cells, 16 bits in ten 5-state cells, and 16 bits in five 7-state cells. 

\textbf{Tradeoff Between Capacity and Lifetime}. The endurance of a flash cell ($L$) has a power-dependence on the number of program/erase (P/E) cyclings ($N_{PE}$)~\cite{Yang:ICSSICT2006,Mohan:FAST2010}, i.e., $L	\propto N_{PE}^\beta$, where $\beta\geq 0.3$. Producing more $V_{th}$ states in a flash cell requires more program operations~\cite{Kim:ISSCC2020}, thereby greatly decreasing the cell endurance. Compared to a 3-bit TLC, a 2-state cell shown in Figure~\ref{f:task_2_all}(a) prolongs the cell endurance by $10\times$~\cite{Gu:ICCD2021}. Although prior work~\cite{Jimenez:DATE2013} improves the flash lifetime by converting 2-bit MLC to 1-bit SLC, the 50\% capacity loss does not maintain a gradual and graceful capacity degradation. In contrast, FRAC can gracefully explore the tradeoff between the chip endurance and capacity, as shown in Figure~\ref{f:task_2_all}(d). By gradually reducing the number of $V_{th}$ states from 8 to 2, FRAC gracefully degrades the capacity of a page from 4KB to 1.3KB and prolongs the cell endurance from $1\times$ to $10\times$.

\textbf{Read and Write}. To sense data from a TLC, as Figure~\ref{f:task_2_all}(e) exhibits, totally $\log_2(8)=3$ iterations are required. The reading reference of this iteration is decided by the result of the previous iteration. For example, after comparing against $r_3$, if the $V_{th}$ state of the cell is larger than $r_3$, $r_5$ will be selected in the next iteration. Otherwise, $r_1$ will be chosen. The same as conventional MLC, TLC, and QLC, a read operation on a $m$-state faction cell also requires $\lceil\log_2(m)\rceil$ iterations. After the data are sensed out of the cells in a page, the data in multiple cells is translated to one value. After an erase operation, to program a TLC, the incremental step pulse scheme~\cite{Suh:JSSC1995} issues multiple programming pulses, each of which is stronger than the previous one, as shown in Figure~\ref{f:task_2_all}(f). Between two programming pulses, there is a verify that checks whether the cell reaches the target $V_{th}$ or not. On the contrary, to program an $m$-state cell, instead of a small pulse, the incremental step pulse scheme can directly start with a larger pulse. In this way, FRAC requires fewer pulses and thus prolongs the cell endurance, when $m < 8$.

\subsection{ESE - An Environmental Sustainability Estimator}
We introduce the Environmental Sustainability Estimator (\textit{ESE}), an innovative tool facilitating accurate assessment of CO2 emissions linked with user activities. ESE integrates environmental sustainability into data center billing policies, incentivizing users to adopt more sustainable computing practices. Illustrated in Figure~\ref{f:task_3_all}(a), ESE comprises a hardware estimator, a data center energy model, and an energy source predictor.
\textit{First}, the hardware estimator analyzes user source code and expected Quality of Service (latency), partitions tasks, and distributes them across CPUs, GPUs, and PIMs to ensure latency requirements are met. It also assesses storage capacity and network bandwidth based on task data access patterns. 
\textit{Second}, the data center energy model will use the hardware estimator output to produce both operational energy consumption ($E_{ope}$) and embodied energy consumption ($E_{emb}$). 
\textit{Third}, the user inputs the task's start time to access weather forecasts. The energy source predictor utilizes this information along with calendar data to determine the net energy demand, representing the disparity between total data center operational energy and current renewable energy generation.
\textit{Finally}, based on the values of $E_{ope}$, $E_{emb}$, and net energy demand, the data center uses different billing policies to decide the user charge.

\textbf{Hardware estimator}.
The hardware estimator minimizes $E_{ope}$ and $E_{emb}$ for each user task within a latency constraint on a heterogeneous hardware platform. It takes source code and expected latency as inputs. The source code is compiled into binaries for an initial hardware configuration, and static features are extracted directly. Runtime features are collected by executing the binaries on real hardware. A latency model processes both types of features to estimate CPU/GPU/PIM/net/SSD latency. The task's estimated latency is the sum of these values. If the estimated latency exceeds the expected latency, kernels are moved to GPUs or PIMs, and the estimator iterates until the expected latency is achieved. Otherwise, the minimum estimated latency is returned.

Traditional task partition methods~\cite{Wang:PACT2010,Wen:HiPC2014} rely on \textit{static features} such as branch count and arithmetic operations, extracted from source code using intermediate representations (IRs). However, these handcrafted features often overlook inter-kernel data flows, treating each kernel independently. Our hardware estimator employs a transformer architecture~\cite{Svyatkovskiy:SECSFSE2020} to directly extract more informative static features from the source code, capturing interactions among all kernels. Leveraging a pre-trained transformer~\cite{Feng:EMNLP2020} tailored for code generation eliminates the need for training from scratch. \textit{Runtime features} encompass CPU cache misses, GPU core utilization, PIM computing unit utilization, and host-device data transfers, among others. Measured using performance profiling tools like Intel VTune and NVIDIA visual profiler, these features are assessed after deploying compiled binaries on real hardware.

To expedite latency evaluation, we train a CNN as a latency model utilizing static and runtime features alongside hardware configuration to predict latency. We create a latency dataset by generating varied partition results and measuring latency values on CPUs, GPUs, and PIMs for training. Post-training, partition results won't require execution on real hardware; instead, latency models will directly provide values.

\textbf{Data Center Energy Model}.
The energy model comprises an operational and an embodied energy model. The energy model inputs are the static and runtime features, hardware configuration, and hardware latency values, while its outputs are operational energy and embodied energy. We will train a CNN as the operational energy model. Similar to latency models in the hardware estimator, we will randomly generate different partition results and measure the energy consumption on CPUs, GPUs, PIMs, networks, and storage systems to build an energy dataset for training the energy model. Besides the dynamic and leakage energy consumption of various hardware units, we will also consider the cooling energy cost, idle equipment energy, power delivery loss, and other operation-related energy in the operational energy model. The embodied energy model is a linear equation. We will compute the total embodied energy of a hardware unit by considering the energy consumption during production and manufacture, transport, use and maintenance, recycling, and other stages in the whole hardware lifetime. The embodied energy ($E_{emb}$) of a user task can be computed as $\sum_{i\in X} TBE_i\times \frac{latency_i}{lifetime_i}$, where $X$ indicates the hardware units used by the task; $TBE_i$ is the total embodied energy of the hardware unit $i$; $latency_i$ means the latency of the hardware unit $i$; and $lifetime_i$ represents the lifetime of the hardware unit $i$.

\begin{figure*}[t]\centering
\begin{minipage}{0.4\textwidth}
    \center{\includegraphics[width=\linewidth]{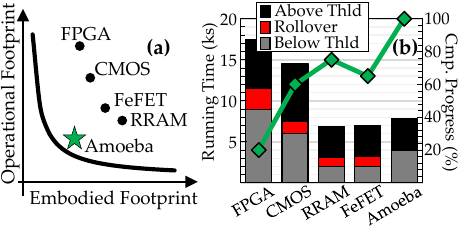}}\vspace{-8pt}
    \caption{Carbon footprints and progresses under energy fluctuations compared to prior designs.}\vspace{-8pt}
    \label{f:task1_results}
\end{minipage}
\hspace{4pt}
\begin {minipage}{0.28\textwidth}
    \center{\includegraphics[width=\linewidth]{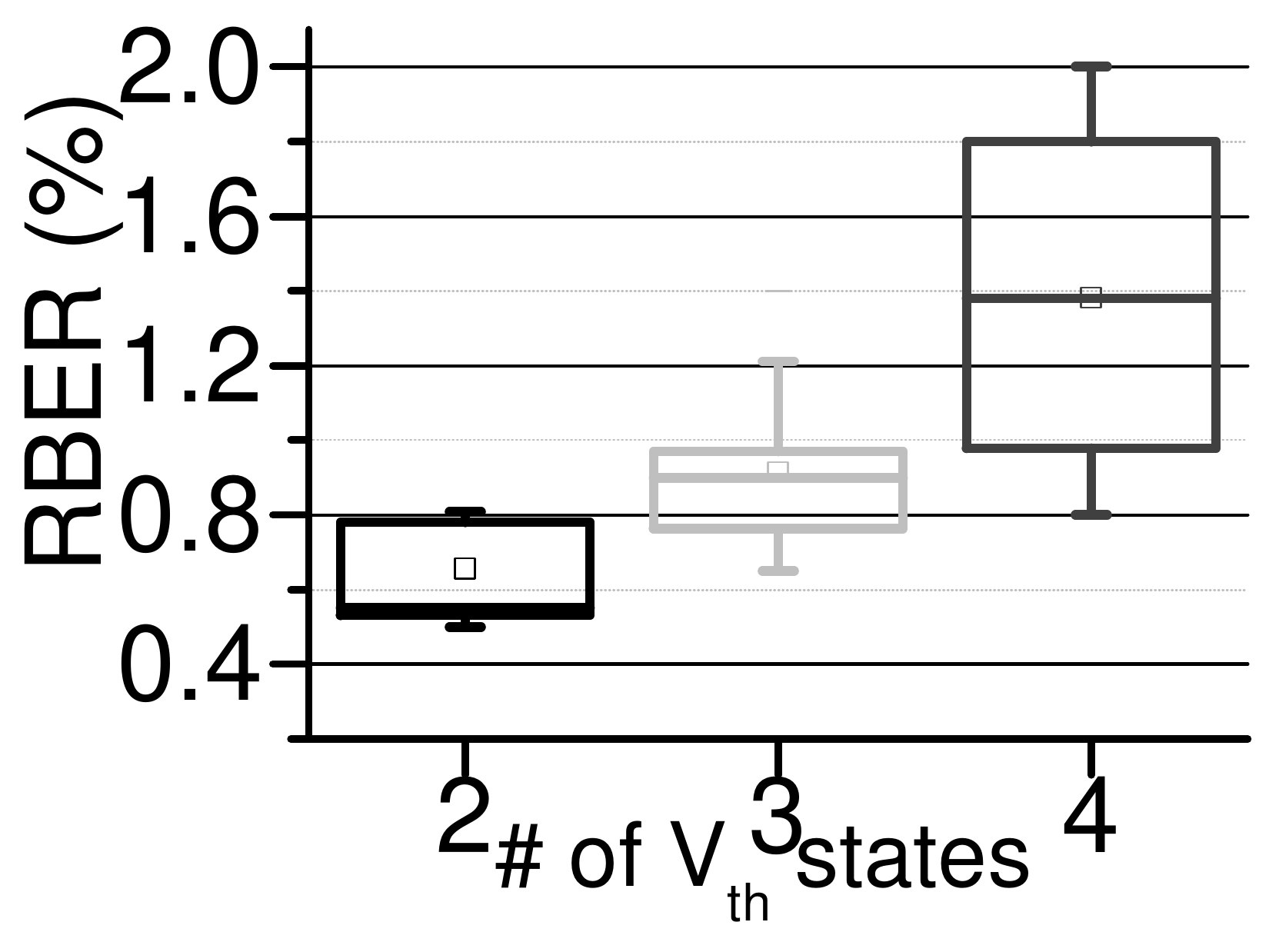}}\vspace{-12pt}
    \caption{The Raw Bit Error Rate (RBER) of FRAC-based pages in a recycled flash chip.}\vspace{-8pt}
    \label{f:co2_pe_expe}
\end{minipage}
\hspace{10pt}
\begin {minipage}{0.26\textwidth}
    \center{\includegraphics[width=\linewidth]{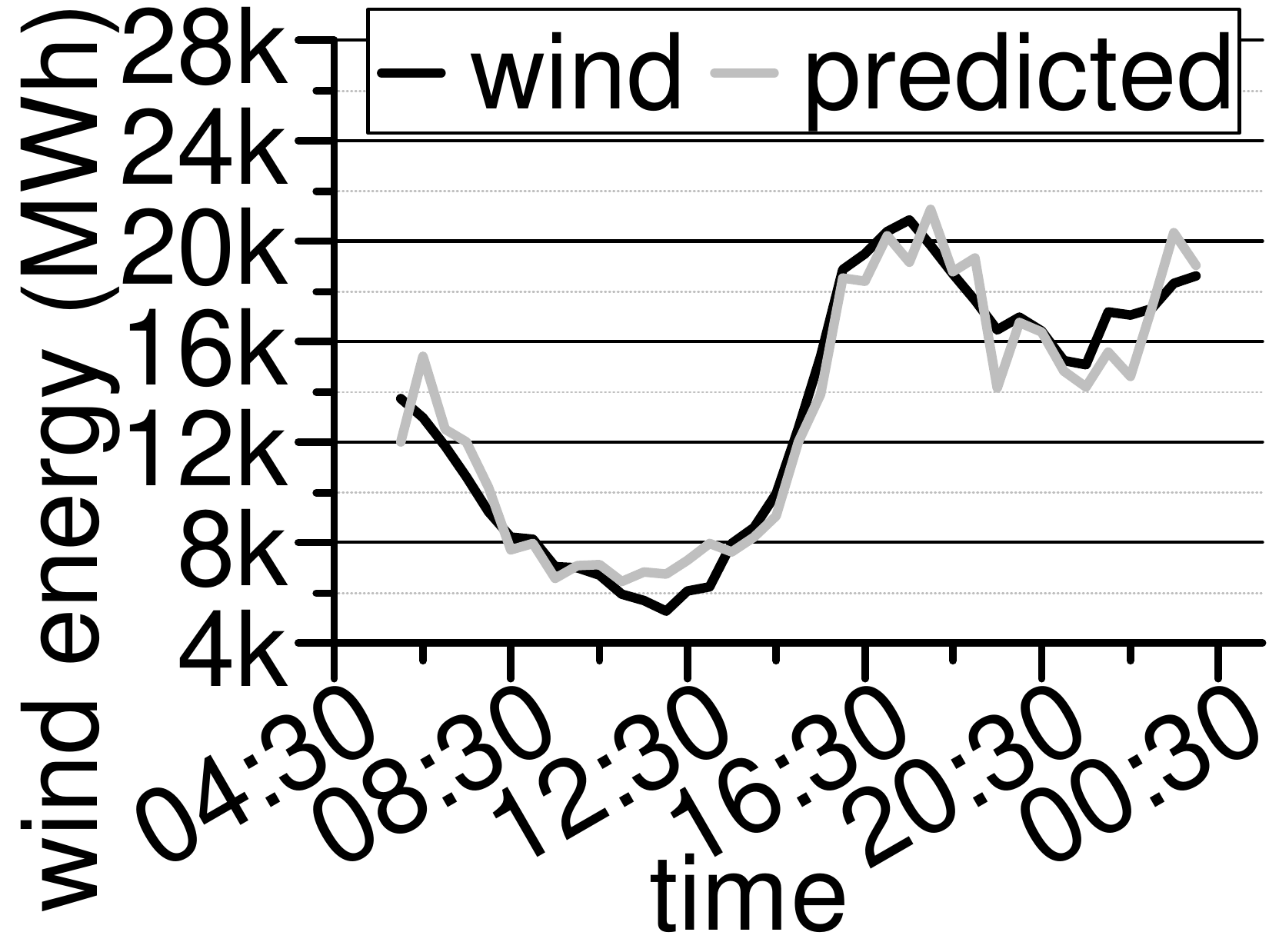}}\vspace{-6pt}
    \caption{The prediction on the wind energy generation in the California grid.}\vspace{-8pt}
    \label{f:co2_weather_pred}
\end{minipage}
\vspace{-10pt}
\end{figure*}

\textbf{Energy Source Predictor}.
The energy source predictor aims to characterize the distribution of net energy demand forecast errors under various weather conditions and data center workload capacities, as shown in Figure~\ref{f:task_3_all}(d). The energy source predictor will be built as a long short-term memory (LSTM) network to output simultaneous quantile forecasts of net energy demand and generated renewable energy. The network inputs include an array of calendar data and weather information. The network will be trained on historical records and predicts the forecasts and forecast errors of $T_0+5$-minute, $T_0+10$-minute, and $T_0+15$-minute at $T_0$. After training, the network will ingest near-past 5-minute forecasts as well as near-past 10-minute and near-past 15-minute historical data available at $T_0$. More specifically, the network predicts the conditional quantiles of forecast errors for net energy demand and renewable energy production. We will consider seven target quantiles, i.e., P2.5, P5, P25, P50, P75, P95, and P97.5.

\vspace{-2pt}
\section{Results}
We offer preliminary results of our ongoing research. These findings highlight the potential of our work to advance environmentally sustainable computing in data centers.

\textbf{\taskOne~ Results}.
We evaluated \taskOne by implementing a simple prototype. We compare~\taskOne~against prior accelerators designs, i.e., FPGA~\cite{zhang2015optimizing}, CMOS~\cite{chen2016eyeriss}, RRAM~\cite{wang2019reram}, and FeFET~\cite{agrawal2022security}. We use three benchmarks including a 32k NTT using Montgomery reduction with a fixed $q$ =12289~\cite{nejatollahi2020cryptopim}, SHA3 with a 1088 blocksize and 1600-bit state sizes, and AlexNet~\cite{chen2016eyeriss}. For embodied carbon emissions, we mainly focused on manufacturing costs and used the total cost for three different domain accelerators as the overall footprint. We consider California grid~\cite{acun2023carbon} as the renewable power supply and adopt its historical data, taking into account dynamic intermittency and fluctuations. We adopt the recently proposed Carbon Explorer framework~\cite{acun2023carbon} to study the Pareto Frontier analysis and report the design solution for each scheme in Figure~\ref{f:task1_results}(left). In general, the~\taskOne~achieves state-of-the-art carbon minimization. 
Figure~\ref{f:task1_results}(right) compares the breakdown of simulation run times and computational progress for AlexNet inference at different times of a week under California grid~\cite{acun2023carbon} supply. For all architectures, \taskOne~achieves the highest forward progress since the fully nonvolatile accelerator can work below threshold power (i.e., below Thld) with reduced operational energy consumption. In addition, we can also see that the fluctuation of the power supply imposes large rollover penalties for nonvolatile processors and CMOS circuitries in existing RRAM and FeFET accelerators.

\textbf{FRAC Results}.
We evaluated the viability of FRAC using a prototype built upon a Zynq FPGA~\cite{eenews:fpga2019} with a NAND flash daughter broad. Recycled flash chips can be plugged into the flash daughter board, while we implement the control logic for FRAC read, program, and erase operations on the FPGA that can access those chips. Our prototype is a minimal storage system built upon recycled flash chips and can be used to study the lifetime improvement achieved by FRAC. Our prototype primarily translates high-level read/write I/O requests to device-level NAND flash commands. We tested the RBER of FRAC cells with different numbers of $V_{th}$ states using our prototype.
Figure~\ref{f:co2_pe_expe} highlights the RBER of FRAC-based pages subjected to 6k programming and erasure cycles in an aged flash chip. A higher RBER indicates a page nearing the end of its lifetime~\cite{Grupp:FAST2012}. When the cells in the recycled flash chip have only two $V_{th}$ states, the RBER is only $0.6\%$ on average. In contrast, three $V_{th}$ states in a cell increase the RBER to $0.9\%$ on average. Four $V_{th}$-state-cells further enlarge the RBER of the same page to 1.4\% on average. These data suggest that a full-fledged FRAC design can fully explore the trade-off between cell endurance and chip capacity.

\textbf{ESE Results}.
We evaluated ESE using a simplified prototype of the energy source predictor. Our prototype predicts the average value of renewable energy generated by wind in the California grid every 30 minutes, utilizing a 2-layer LSTM network with forget, input, and output gates. 
Historical wind energy generation data from the California grid served as ground truth, sourced from California ISO~\cite{California:CISO2021}, while wind data from Apple Weather~\cite{Dark:WEATH2021} provided input.
Training utilized 70\% of the data, with 10\% for validation and the remaining 20\% for testing. Figure~\ref{f:co2_weather_pred} illustrates the average wind energy generation prediction. While our prototype's predictions capture the general trend of the ground truth data, the 30-minute prediction interval results in larger fluctuations compared to the ground truth, suggesting the need for shorter intervals (e.g., $5\sim15$ minutes). These findings imply that a comprehensive ESE could accurately evaluate operational and embodied energy consumption in data centers.

\section{Conclusion}
Modern data centers face significant environmental sustainability challenges throughout their lifecycle. The project aims to tackle these challenges by developing novel technologies including a reconfigurable hardware accelerator architecture, a fraction NAND flash cell and its system integration, and an environmental sustainability estimator. These advancements lay a solid foundation for improving environmental sustainability in data centers, enabling both users and data centers to conduct data-intensive applications in a more sustainable and scalable manner. This not only enhances computing capabilities but also maximizes the societal benefits derived from computing.


\bibliographystyle{ieeetr}
\bibliography{reference}
\end{document}